\documentclass[12pt,preprint]{aastex} % for a referee version
\usepackage[]{natbib}
\shorttitle{Kormendy relation up to $\mathrm{z \sim 0.64}$}
\shortauthors{La Barbera et al.}

\def\coma{$\mathrm{Coma}$}
\def\a209{$\mathrm{A\,209}$} 
\def\ac118{$\mathrm{AC\,118}$}
\def\eis0048{$\mathrm{EIS\,0048}$} 
\def\re{$\mathrm{r_e}$}
\def\RE{$\mathrm{R_e}$}
\def\s0{$\mathrm{\sigma_0}$}
\def\lre{$\mathrm{\log r_e}$}
\def\lRE{$\mathrm{\log R_e}$}
\def\mie{$\mathrm{< \! \mu \! >_e}$} 
\def\dlre{$\mathrm{\delta(\log r_e)}$}
\def\dmie{$\mathrm{\delta( < \! \mu \! >_e )}$} 

\def\ns{$\mathrm{n}$}
\def\mlre{$\mathrm{MLS_{\log r_e}}$}
\def\mmie{$\mathrm{MLS_{<\!\mu\!>_e}}$}
\def\bmls{$\mathrm{BMLS}$}
\def\smie{$\mathrm{\sigma^{(i)}_{<\!\mu\!>_e} }$}

\begin{document}

\title{On the Invariant Distribution of Galaxies in the \re\,--\mie\,
plane out to $\mathrm{z = 0.64}$\footnote{Based on observations
collected at European Southern Observatory} }

\author{F. La Barbera, G. Busarello, P. Merluzzi, M. Massarotti}
\and
\author{M. Capaccioli}
\email{labarber@na.astro.it}
\affil{INAF Osservatorio Astronomico di Capodimonte, Napoli, Italy}

\begin{abstract}
We study the evolution of the relation between half-light (effective)
radius, \re, and mean surface brightness, \mie, known as Kormendy
relation, out to redshift $\mathrm{z=0.64}$ in V-band restframe on the
basis of a large sample of spheroidal galaxies ($\mathrm{N=228}$)
belonging to three clusters of galaxies.  The present sample
constitutes the largest data set for which the Kormendy relation is
investigated up to a look-back time of $\mathrm{\sim 6~Gyr }$ ($ \rm
H_0= 70~Km s^{-1} Mpc^{-1}, \Omega_M=0.3, \Omega_{\Lambda}=0.7$). A
new fitting procedure, which suitably accounts for selection criteria
effects, allows for the first time to study the trend of the slope
($\beta$) and of the intrinsic dispersion (\smie) of the Kormendy
relation, and the properties of the whole distribution in the \re \,
-- \mie \, plane as a function of look-back time.  The slope $\beta$
of the relation does not change from $\mathrm{z=0.64}$ to the present
epoch: $\beta=2.92\pm0.08$, implying a tight constraint of
$18$--$28\%$ on the variation of the stellar formation epoch along the
sequence of spheroidal galaxies per decade of radius. The intrinsic
dispersion of the relation, $\mathrm{\sigma^{(i)}_{<\!\mu\!>_e}
=0.40\pm0.03}$, does not vary with redshift and the distribution of
galaxy sizes as well as the distribution in the plane of the effective
parameters do not vary among the clusters, as proven by the
Kolmogorov--Smirnov tests.  We conclude that, whatever the mechanism
driving galaxy evolution is, it does not affect significantly the
properties of bright galaxies in the \lre \, -- \mie \, plane at least
since $\mathrm{z=0.64}$.  The evolution of the zeropoint of the
Kormendy relation is fully explained by the cosmological dimming in an
expanding universe plus the passive luminosity evolution of stellar
populations with high formation redshift ($\mathrm{z_f>2}$).

\end{abstract}
%
%________________________________________________________________

\section{Introduction}
Early--type galaxies (ETGs) are known to populate a two-dimensional
manifold in the space of their observed quantities. This Fundamental
Plane (FP) is usually expressed by a correlation among the effective
radius \re, the mean surface brightness \mie \, within \re, and the
central velocity dispersion \s0:
\begin{equation}
\mathrm{
\log r_e = a \cdot \log \sigma_0 + b \cdot <\!\mu\!>_e + c. 
} \label{FPEQ}
\end{equation}
Due to the small intrinsic dispersion ($\sim 10 \%$ in \re \, and \s0,
and $\mathrm{\sim 0.1~mag}$ in \mie), the FP is considered as a
powerful tool to measure galaxy distances and to constrain the
processes driving galaxy evolution (e.g. \citealt{KJM93, JFK96, JOR99,
KIV00}).

A relevant projection of the FP is the correlation between \RE \, and
\mie, also known as Kormendy relation (hereafter KR):
\begin{equation}
\mathrm{
 <\!\mu\!>_e=\alpha + \beta \cdot \log R_e,   \label{EQKR}
}
\end{equation}
where \RE \, is the effective radius in $\mathrm{kpc}$, and $\beta
\simeq 3$ \citep{KOR77}. Different studies have demonstrated that in
the nearby universe bright ETGs define a sharp sequence in the
\lRE--\mie \, plane, with an intrinsic dispersion of $\sim 0.3$ --
$0.4$ in \mie \, (\citealt{HaK87, HOS87, SaP91, SaL01}). Moreover,
\citet{CCD92} have shown that $\mathrm{ETGs+bulges}$ form two distinct
families in the plane of the effective parameters: that of the bright
ETGs, following the KR, and another `ordinary' family, whose
properties are more disperse and heterogeneous. Recently, \citet[see
also references therein]{GrG03} demonstrated that the claimed dichotomy
between the bright and the dwarf ETGs is only apparent and that there
is a continuous structural relation between the two classes. In
particular, the different behavior of bright and dwarf ETGs in the
\lRE--\mie \, plane, and the change in slope of their relations, do
not imply a different formation mechanism but can be interpreted with
systematic changes in the profile shape with galaxy magnitude.

In order to understand the processes underlying galaxy formation and
evolution, it is crucial to investigate the FP and the other relations
between galaxy parameters at different look-back times. In the
monolithic formation scenario \citep{Lar74}, ETGs form at very high
redshifts in a strong burst of star formation, and the distribution of
galaxies in the space of observed quantities changes with redshift due
to the passive fading of their stars.  In the hierarchical framework,
bright ETGs form by the progressive merging of smaller units, and the
FP relations are built up with redshift.  Different works have shown
(1) that (dissipation-less) merging seems to move galaxies along the
FP \citep{CdCC95, DCR02}, and (2) that in hierarchical models, the
luminosity--weighted ages of stellar populations in cluster galaxies
appear to be the same of those predicted for pure passive evolution
(e.g. \citealt{KAU96}).  All these facts suggest that for both
scenarios dynamical effects could play a minor role in the evolution
of scaling correlations. It is expected that the FP slopes change with
redshift mainly for possible differences in properties of galaxy
stellar populations (such as age and metallicity) along the early-type
sequence, while the zero point varies consistently with the evolution
of a passive stellar population.  In the hierarchical framework,
however, it is also expected that the distribution in the space of
observed quantities evolves with redshift, as merging proceeds to
larger and larger sizes. Clearly, in order to analyze this subject,
large samples of galaxies are needed at different look-back times.

In the recent years, it has become feasible to study the scaling
relations at intermediate redshifts. Due to the heavy demand of
observing time in velocity dispersion measurements, many studies have
investigated pure photometric laws, like the KR. The main effort to
constrain the evolution of the KR slope with redshift has been
performed by \citet{ZSB99}, who analyzed ETGs properties in four
clusters at $\mathrm{z \sim 0.4}$ and in one at $\mathrm{z=0.55}$,
with an average of $\sim 20$--$30$ galaxies per cluster. These authors
found that the slope is comprised between 2.2 and 3.5 both for the
Coma and for the distant clusters, depending on the selection criteria
of the samples, and suggested, therefore, that it could not have
evolved since $\mathrm{z \sim 0.55}$.

The evolution of the KR zero point for cluster galaxies has been
investigated in different works in order to perform the Tolman surface
brightness test for the universal expansion \citep{SaP91, PDdC96,
SaL01, LuS01a, LuS01b, LuS01c} and to derive the luminosity evolution
of ETGs (e.g. \citealt{BAS98, ZSB99}). All these studies concluded
that the measured change is consistent with what expected on the basis
of cosmological dimming plus passive aging of stellar populations
formed at high redshifts.

The KR was investigated at high redshifts for galaxies in the Hubble
Deep Field by \citet{FCA98}, who found that ETGs in the field seem to
follow the same restframe KR also at redshifts $\mathrm{z \sim 2 \! -
\! 3}$.

As shown by \citet{LBM02}, it is possible to derive structural
parameters of galaxies at intermediate redshifts by using
ground--based imaging.  In the present work, we use structural
parameters in the V-band restframe for a total of $\mathrm{N = 228}$
spheroids in the clusters \a209 \, at $\mathrm{z=0.21}$, \ac118 \, at
$\mathrm{z=0.31}$, and $\mathrm{EIS \, 0048 \! - \! 2942}$ (hereafter
\eis0048) \, at $\mathrm{z=0.64}$, and investigate the evolution of
the distribution in the \lre -- \mie \, plane up to a look-back time
of $\mathrm{\mathrm{\sim 6~Gyr}}$. Surface photometry for galaxies in
\a209, \ac118 \, and \, \eis0048 is published in \citet{LBM03b}. 

One of the main challenges of previous works on the subject has been
represented by selection effects, which can critically affect the
determination of the KR.  To work out this problem, we introduce a new
fitting procedure, which provides un-biased estimates of the KR
coefficients, and allows for the first time to investigate the
evolution over redshift of the whole distribution in the plane of
effective parameters.

The layout of the paper is the following. In Section~\ref{SDATA}, we
present the samples used for the analysis. Section~\ref{SKREL}
describes the new fitting procedure and deals with the evolution of
the KR slope and the comparison of the \lre--\mie \, distributions at
different redshifts. Section~\ref{SLEV} shows the luminosity evolution
inferred by the KR. Conclusions are drawn in section~\ref{SCONC}.  In
the following, we will assume the cosmology $\Omega_m=0.3$,
$\Omega_\Lambda=0.7$ and $\mathrm{ H_0= 70~Km s^{-1} Mpc^{-1} }$. With
these parameters, the age of the universe is $\mathrm{\sim13.5~Gyr}$,
and the redshifts $\mathrm{z=0.21, z=0.31}$, and $\mathrm{z=0.64}$
correspond to look-back times of $\mathrm{\sim 2.5, 3.5}$, and
$\mathrm{6~Gyr}$ respectively.

\section{The samples}
\label{SDATA}
Photometry for the three clusters of galaxies \a209 \,
($\mathrm{z=0.21}$), \ac118 \, ($\mathrm{z=0.31}$), and \eis0048 \,
($\mathrm{z=0.64}$) resides on data collected at the ESO New
Technology Telescope (NTT) and at the ESO Very Large Telescope
(VLT). The data relevant for the present study include images in the
R band for \a209 \, and \ac118, and in the I band for \eis0048, all
corresponding approximately to V-band restframe. For \ac118 \, and
\eis0048, galaxies were selected according to the photometric redshift
technique (see~\citealt{MIB01}, and references therein), while for
\a209 we considered the sample of spectroscopically confirmed cluster
members from \citet{MGB03a} complemented with photometric data (see
below).  The relevant information on the samples are summarized in
table~\ref{DATA}.

Structural parameters were derived as described in \citet{LBM02}, by
fitting galaxy images with Sersic models convolved with suitable
representations of the PSF. Uncertainties on structural parameters
were estimated as described in \citet{LBM03b}, and typically amount to
$\delta(\log r_e)\sim0.09$ and $\delta(<\!\mu\!>_e)\sim0.4$ for \a209,
$\delta(\log r_e)\sim0.14$ and $\delta(<\!\mu\!>_e)\sim0.55$ for
\ac118, and to $\delta(\log r_e)\sim0.07$ and
$\delta(<\!\mu\!>_e)\sim0.30$ for \eis0048. In order to study the KR,
we selected the population of spheroids on the basis of the shape of
the light profile, as parametrized by the Sersic index \ns. We
classified as spheroids the galaxies with $\mathrm{n} > 2$,
corresponding to objects with a bulge fraction greater than $\sim20\%$
(see~\citealt{SAG97, vDF98a}). For each cluster we considered all the
galaxies brighter than a given magnitude limit $\mathrm{m_c}$. Few
objects with small radii and poorly determined structural parameters
were excluded by selecting only galaxies with \mie \, greater than a
given surface brightness limit $\mathrm{<\!\mu\!>_{e,c}}$.  The
completeness of the spectroscopic sample of \a209 \, is about $ 50\%$
at $\mathrm{R\sim20}$, and falls to zero at $\mathrm{R=21}$.  A
magnitude complete sample was obtained by adding to the spectroscopic
data all the spheroids within the colour--magnitude sequence of \a209
(see~\citealt{MGB03a}) down to $\mathrm{R\sim20.1}$.  We point out
that the present samples constitute the largest data set ($\mathrm{N
= 228}$) of cluster galaxies, for which the Kormendy relation is
investigated up to a look-back time of $\mathrm{\sim6~Gyr}$.

In order to compare the distributions in the \lre--\mie \, plane at
intermediate redshifts with that of nearby galaxies, we took advantage
of the large sample of ETGs in the Coma cluster by \citet[hereafter
JFK95a]{JFK95a}. Structural parameters for $\mathrm{N=147}$ galaxies
were taken in JFK95a, who fitted the growth curve of aperture
magnitudes with de Vaucouleurs $\mathrm{r^{1/4}}$ profiles and
corrected the estimated parameters for seeing effects. The typical
uncertainties on \lre \, and \mie \, amount to $\sim0.08$ and
$\mathrm{\sim0.3~mag}$, respectively (see table~11 of JFK95a).  Since
it was not possible to select galaxies in the JFK95a sample by using
Sersic indices as for the other clusters, we adopted the relation
between Sersic index and central velocity dispersion recently found
for nearby galaxies by \citet{GRA02}: $\log n \propto \log
\sigma_0$. We considered the sample of Coma ETGs with available
velocity dispersion from \citet[hereafter JFK95b,
$\mathrm{N=76}$]{JFK95b} and from \citet[hereafter J99,
$\mathrm{N=116}$]{J99}. Among these galaxies, $\mathrm{N=113}$ objects
have surface photometry from JFK95a. According to \citealt{GRA02}
(Figure~1), we selected galaxies with $\mathrm{\sigma_0 > 110~km \,
s^{-1}}$, which corresponds to $\mathrm{n \gtrsim 2}$, resulting in a
sample of $\mathrm{N=93}$ objects. We also excluded from the
calculation of the fitting coefficients the three galaxies with
largest size in the JFK95a sample: due to the presence of extended
halos the use of a de Vaucouleurs model can give very different
parameters with respect to the estimates obtained by the Sersic
models.  The final sample consists in $\mathrm{N=90}$ galaxies
brighter than $\mathrm{r_G\sim15.3}$, where $\mathrm{r_G}$ is the
Thuan--Gunn r-band magnitude~\footnote{Since spheroids have very small
internal colour gradients at optical wavelengths, the difference
between the $\mathrm{r_G}$ band and the V-band restframe at the \coma
\, redshift is not relevant for the present study}. We note that, due
to the previous selections, the Coma sample does not constitute a
magnitude complete sample of galaxies. This point will be further
addressed in Section~\ref{FITKR}.

For \a209, \ac118, and \eis0048, mean surface brightnesses were
corrected for galactic extinction following \citet{SFD98}, while the
cosmological dimming was not removed. For the \coma \, sample, since
both corrections were applied by JFK95a, we added the term $10 \cdot
\log(1+z)$ to the \mie \, values.
%
% per EIS0048  0.049 ---> 0.03
% per AC118    0.035 ---> 0.01 stesso problema di AC118
% per A209     0.051  ??? valore negativo di E(B-V)
% per Coma     0.0207 --> 0.02590

\section{Kormendy relations}
\label{SKREL}
\begin{figure}
\plotone{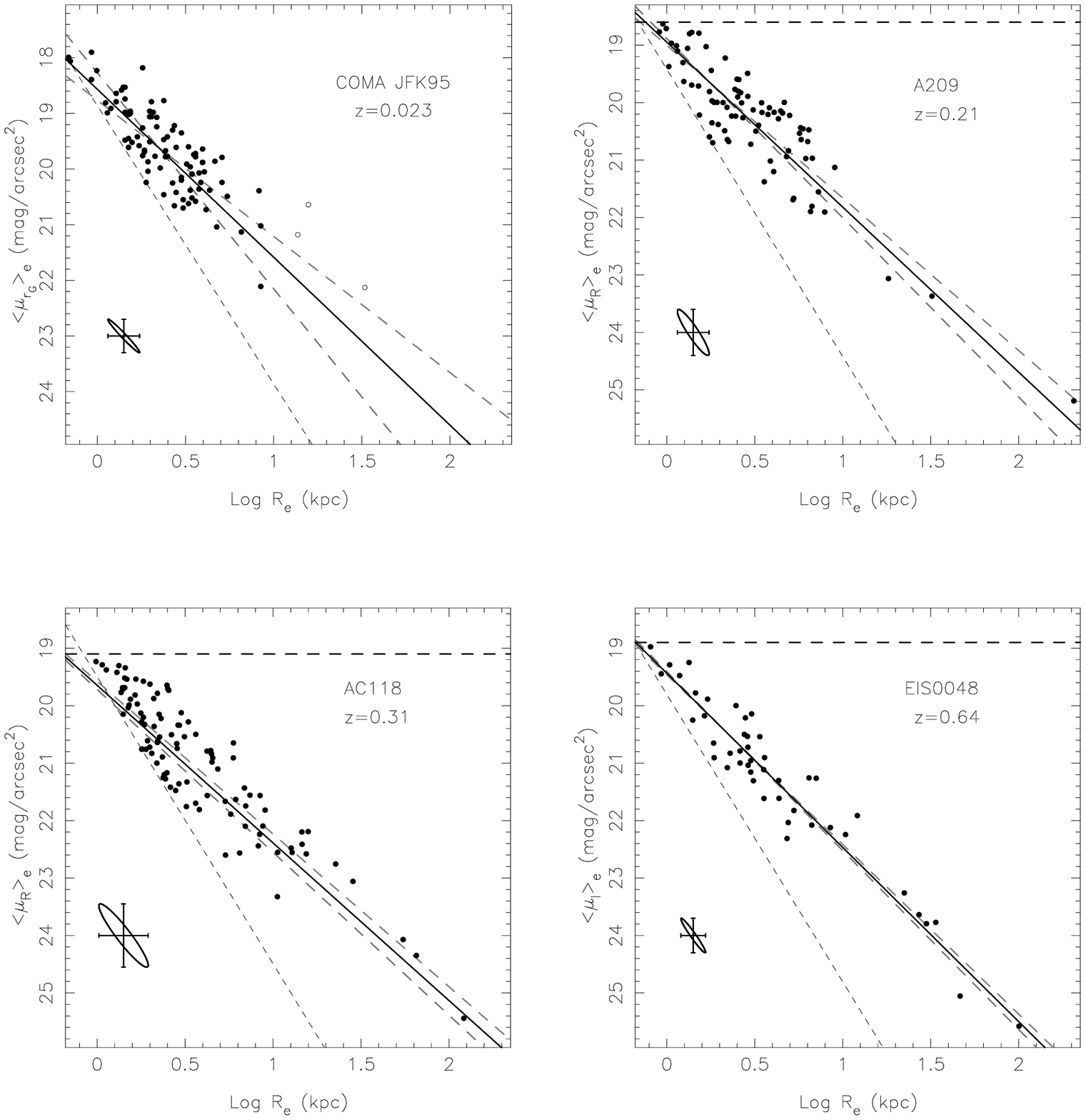}
\caption{Kormendy relations for Coma, \a209, \ac118, and \eis0048.
The range of \lRE \, is the same for each panel, while the range of
\mie \, is the same for \a209, \ac118, and \eis0048, and is shifted by
$\mathrm{-1~mag \, arcsec^{-2}}$ for Coma. The \mlre \, and \mmie \,
fits are represented by the long--dashed lines, while the solid line
is the \bmls \, regression. The short--dashed line and the horizontal
line indicate the cuts in magnitude and \mie \, for each sample. The
correlation of the uncertainties on \lre \, and \mie \, are shown by
the ellipses ($1~\sigma$ confidence contours) in the lower--right of
each panel. The galaxies that were not considered in the fits for the
\coma \, cluster are marked by empty circles.}
% BoundingBox 0 170 586 752
\label{KREL}
\end{figure}

The \lRE--\mie \, diagrams for \coma, \a209, \ac118, and \eis0048 \,
are shown in Figure~\ref{KREL}. For each cluster, spheroids follow a
sharp, well defined KR.

\subsection{Fitting the KR}
\label{FITKR}
Since selection effects can strongly affect the estimate of the KR
coefficients (see~\citealt{ZSB99}), we fitted the \lre--\mie \,
sequences by introducing a modified least square procedure (hereafter
$\mathrm{MLS}$, see appendix~\ref{appA}), which corrects the bias due
to the different completeness cuts in magnitude and \mie \, of each
sample.  The fitting coefficients were derived by minimizing the rms
of the residuals with respect to \lre \, and \mie \, (\mlre \, and
\mmie \, fits), and by applying the bisector regression (\bmls,
see~\citealt{AkB96}), which is more stable and effective with respect
to the other fitting methods.

\begin{figure}
\epsscale{0.6}
\plotone{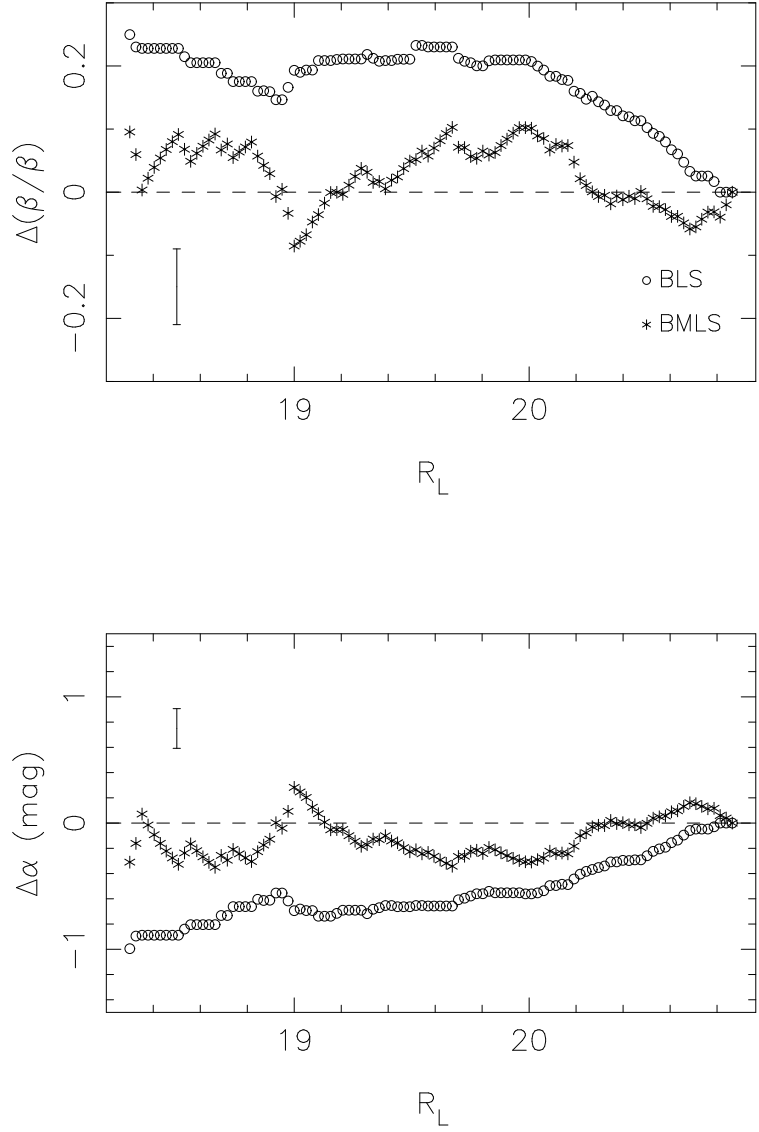}
\caption{Bias on the KR coefficients of \ac118 as a function of the
magnitude cut, $\mathrm{R_L}$. Different symbols (see the lower--right
corner of the upper panel) denote the bisector ordinary least square
(BLS) and the bisector modified least square (BMLS) fits,
respectively. In each panel, the mean error bars on the coefficients
are shown for reference.}
\label{BIAS_FITKR}
\end{figure}

In order to illustrate the reliability of the adopted procedure, we
applied different magnitude cuts, $\mathrm{R_L}$, to the sample of
\ac118, which has the largest number of galaxies, and for each value
of $\mathrm{R_L}$ we derived the slope and the zero point of the KR by
using the $\mathrm{MLS}$ and the ordinary least square
($\mathrm{OLS}$) fits.  In Figure~\ref{BIAS_FITKR} we show the
variations of $\beta$ and $\alpha$ as a function of $\mathrm{R_L}$ for
the bisector fit. Similar results are obtained for the \lre \, and the
\mie \, regressions. The completeness cut affects significantly the
OLS method, producing the well known Malmquist bias: for lower values
of $\mathrm{R_L}$ the slope of the KR is overestimated while the value
of $\alpha$ decreases systematically. On the contrary, it is evident
that the $\mathrm{MLS}$ procedure removes efficiently the systematic
trends of the $\mathrm{OLS}$ method, providing un-biased estimates of
the KR coefficients.  This implies that the $\mathrm{MLS}$ fit allows
a straightforward comparison of the slope and of the zeropoint of the
KR, ruling out any effect related to the different selection criteria
of the samples.

Since the selection criteria for the Coma cluster are different with
respect to those of the other clusters, we also verified that the
fitting results for this sample are not affected by the cuts in
magnitude and velocity dispersion.  According to the scatter of the
relation between velocity dispersion and Sersic index by
\citet{GRA02}\footnote{We assumed a $2 \sigma$ dispersion of $\mathrm{
\sim 50 \, km \, s^{-1} }$}, we varied the velocity dispersion cut
within $\mathrm{\pm 25 \, km \, s^{-1}}$ and re-derived the
$\mathrm{MLS}$ coefficients.  It turned out that the slope $\beta$ of
the KR changes by few percents, while the corresponding variations of
the zeropoint are within $\mathrm{ \pm 0.1 mag }$.  These values are
fully consistent with the estimates of $\beta$ and $\alpha$ that are
given in Sections~\ref{SBEV} and~\ref{SLEV}. A similar result was
obtained by changing the magnitude limit of the sample down to
$\mathrm{r_g=14.5}$. It is worth to be noted that the values we
obtained for the coefficients of the KR at $\mathrm{z \sim 0}$ are in
very good agreement (see Sections~\ref{SBEV} and~\ref{SLEV}) with
those of previous works for bright nearby ETGs, spanning a much wider
luminosity range (see e.g.~\citealt{CCD92}). The selection effects on
the local reference sample do not affect, therefore, the present
analysis.

Another important point is the correlation between the uncertainties
on galaxy parameters: for the effective parameters, we have
$\mathrm{\delta(\log r_e)= \gamma \cdot \delta( < \! \mu \! >_e)}$
(see~JFK95a), where the value of $\gamma$ depends on the methods used
to derive \re \, and \mie, and the corresponding uncertainties. In
order to estimate the range of possible values of $\gamma$, we
compared the uncertainties on \lre \, and \mie \, for samples of
galaxies from different sources (\citealt{LBE91, SBD93, JFK95a, SGH97,
SGB98, MGA99}): we found $0.25 \lesssim \gamma \lesssim 0.34$. By
using numerical simulations (see~\citealt{LBM02}) and by comparing
repeated measurements of \re \, and \mie \, (see~\citealt{LBM03b}), we
obtained $\gamma \sim 0.25$ for the parameters of \a209, \ac118, and
\eis0048, in agreement with the quoted range of $\gamma$ values.  The
correlation between \dlre \, and \dmie \, and the typical values of
the uncertainties on the effective parameters are shown in
Figure~\ref{KREL}. The effect on the KR coefficients was estimated by
using numerical simulations of the \lre--\mie \, diagrams, by a
procedure analogous to that described in \citet{LBC01}.  We found that
the bias on the KR coefficients is negligible, amounting at most to
$\sim 3\%$ for \ac118. This is in agreement with what found by
\citet{HaK87} for the \lre--\mie \, relation of nearby bright
ellipticals and spiral bulges.

\subsection{Evolution of the slope}
\label{SBEV}
\begin{figure}
\epsscale{0.8}
\plotone{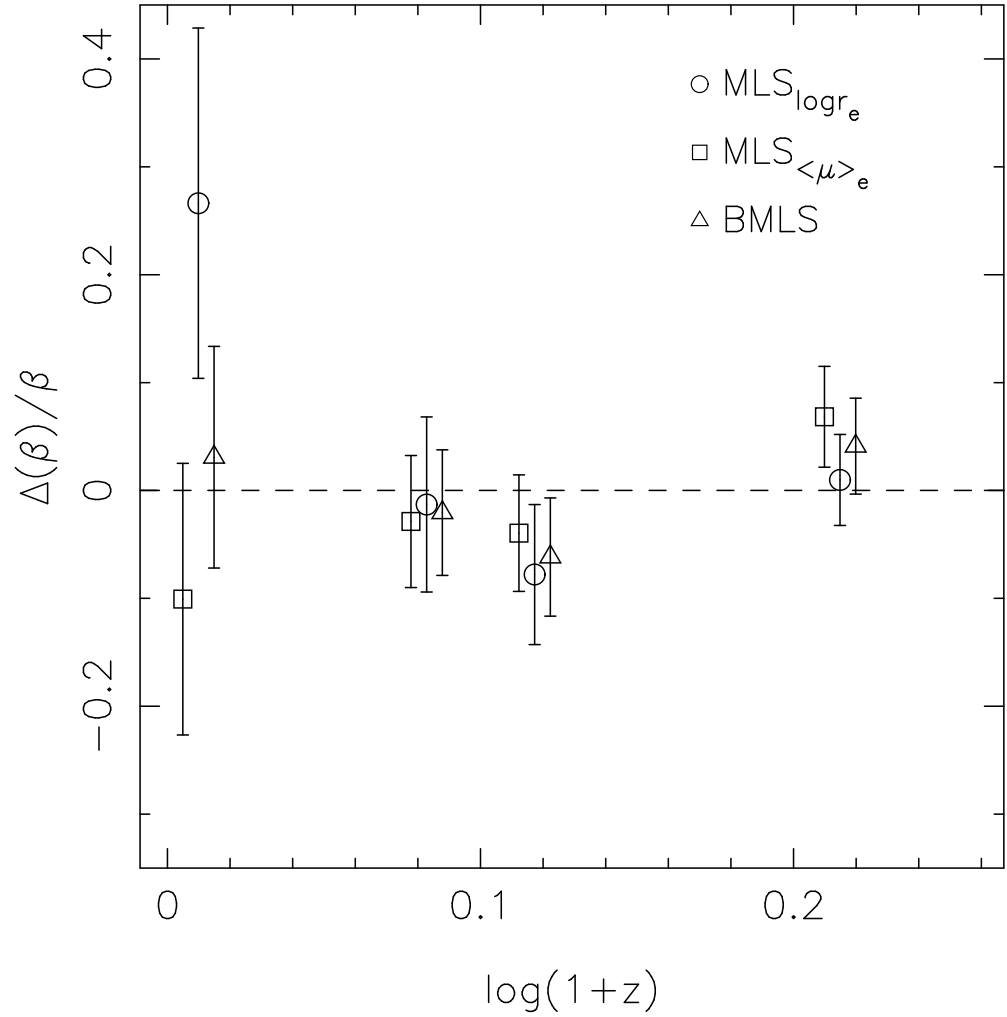}
\caption{Evolution with redshift of the relative slope of the KR. The
symbols denote the different fitting methods (see text). For each
fitting method, the relative variations were computed with respect to
the mean values of table~\ref{KRMSLOPE}.
\label{BEV}
}
\end{figure}

The slopes of the KRs are reported in table~\ref{KRSLOPE}, while the
variation of $\beta$ with redshift is shown in Figure~\ref{BEV}. The
uncertainties were estimated by the bootstrap method and amount to
$\sim4$--$8\%$ for \a209, \ac118, and \eis0048. We note that the
uncertainty for the \coma \, sample is significantly higher ($\sim 9$
-- $12 \%$) due to the smaller range of \lre.  Figure~\ref{BEV}
clearly shows that the values of $\beta$ are fully consistent for each
pair of clusters, except for the \mlre \, and \mmie \, slopes of the
\coma \, sample, which are consistent with the other values at a lower
significance level ($\sim1.5~\sigma$). This is due to the fact that
the \mlre \, and \mmie \, methods are more sensitive, with respect to
the \bmls \, method, to the range of \lre \, used in the
fit\footnote{In fact, if we include also the three largest galaxies in
the fit of the \coma \, sample, the bisector method gives
$\beta=2.6\pm0.4$, that is fully consistent with the value in
table~\ref{BEV}.}.  We conclude, therefore, that no significant
evolution in the slope of the KR is found up to $\mathrm{z\sim 0.64}$.
We also note that the weighted means of $\beta$ reported in
table~\ref{KRMSLOPE} for the various fitting procedures are in very
good agreement with values found by previous studies at $\mathrm{z
\sim 0}$: e.g. $\beta \sim 2.94$ (\citealt{HaK87}), $\beta=3.14 \pm
0.09$ (\citealt{HOS87}) and $ 2.9 \lesssim \beta \lesssim 3.4$
(\citealt{SaL01}).

The evolution of the KR slope with redshift, $\Delta(\beta)$, carries
information on the variation in the properties of the mean stellar
populations (SPs) as a function of galaxy radius:
\begin{equation}
\mathrm{ \Delta(\beta)= \Delta\left(\frac{d \! <\!\mu\!>_e}{d \! \log
 r_e}\right)= \frac{d \! \Delta(<\!\mu\!>_e)}{d \! \log r_e} = \frac{d
 \! E}{d \! \log r_e}, } \label{delb}
\end{equation}
where $\mathrm{E=E(r_e)}$ is the luminosity evolution between
$\mathrm{z=0.64}$ and $\mathrm{z=0}$ at a given \re.  In
appendix~\ref{appB}, we use the GISSEL98 synthesis code
(\citealt{BrC93}) to derive an approximated analytic expression for
$\mathrm{E}$ as a function of the age $\mathrm{t_f}$ and the
metallicity $\mathrm{Z}$ of the galaxy SPs. We consider luminosity
evolution in the V band, since (1) the filters relative to each sample
approximate the V-band restframe and (2) internal colour gradients of
spheroids are known to be very small in the optical wavebands and not
to evolve significantly at intermediate redshifts
(see~\citealt{LBM03b}). For models with formation redshift
$\mathrm{z_f \in [1.8,10]}$ and metallicity $\mathrm{Z \in
[0.2,2.5]Z_\odot}$, which are suitable to describe the population of
spheroids (see also section~\ref{SLEV}), we show that $\mathrm{E}$
depends mainly on $\mathrm{t_f}$, i.e. $\mathrm{E \simeq E(t_f)}$, so
that Eq.~\ref{delb} can be approximated as:
\begin{equation}
\mathrm{ \Delta(\beta) \simeq \frac{d \! E}{d \! \log t_f} \cdot
 \frac{d \! \log t_f}{d \! \log r_e} = F(t_f) \cdot \frac{d \! \log
 t_f}{d \! \log r_e}, }
\end{equation}
where $\mathrm{F(t_f) = d \! E/d \! \log t_f}$ is computed as
described in appendix~\ref{appB}. Assuming that the maximum allowed
variation of $\beta$ is $2~\sigma_\beta$ (see table~\ref{KRMSLOPE}), we
obtain the relation
\begin{equation}
\mathrm{\mid \! d \! \log t_f/d \! \log r_e \! \mid < 2\sigma_\beta /
F(t_f)}.
\end{equation}
For a SSP, $\mathrm{F(t_f)} \in \mathrm{[1.3,1.65]~mag/Gyr}$, and the
value of $\mathrm{2~\sigma_\beta / F(t_f)}$ turns out to be in the
range $[0.1,0.12]$.  This implies that the absolute relative variation
of $\mathrm{t_f}$ per decade of galaxy radius must be smaller than
$\sim 28 \%$ ($\mathrm{\sim0.12~dex}$). As shown in
appendix~\ref{appB}, a lower limit is obtained for SP models with a
protracted SFR. For an exponential SFR with time scale $\mathrm{\tau =
1~Gyr}$, we find $\mathrm{F(t_f)} \in \mathrm{[2,3]~mag/Gyr}$ and the
relative variation of $\mathrm{t_f}$ is constrained to be smaller than
$\sim 18 \%$.

\subsection{Evolution of the intrinsic dispersion}
\label{SSEV}
Since the values of $\beta$ are the same at all redshifts, we adopted
as common slope in the KR fit the value relative to the \bmls \,
regression, $\beta = 2.92 \pm 0.08$, and computed for each sample the
zeropoint $\alpha$ and the scatter $\mathrm{\sigma_{< \! \mu \!
>_e}}$, defined by the rms of the residuals $\delta$ in \mie:
$\mathrm{ \delta = < \! \mu \!  >_e - \alpha - \beta \cdot \log r_e}$.
The uncertainties on $\alpha$ and $\mathrm{\sigma_{< \! \mu \!  >_e}}$
were obtained by numerical simulations, taking also into account the
measurement error on the adopted value of the slope.  We estimated the
intrinsic dispersion of the relations (\smie) by subtracting in
quadrature to $\mathrm{\sigma_{< \! \mu \! >_e}}$ the amount of
dispersion on $\delta$ due to the measurement errors on \mie \, and
\lre. To this aim, we took into account the typical uncertainties on
\lre \, and \mie \, (see section~\ref{SDATA}), and their
correlation\footnote{The covariance term between the measurement
errors on \lre \, and \mie was estimated by the same procedure adopted
for the typical uncertainties on structural parameters (see
section~\ref{SDATA}). For the \coma \, cluster, we adopted the
covariance term used for \eis0048, whose structural parameters have
similar uncertainties.}.  The values are given in
table~\ref{KRPAR}. It is clear from the table that the KR has
significant intrinsic dispersion also at intermediate redshifts
($\mathrm{\sim 0.4}$ in \mie), and that this dispersion does not
change with redshift.  The intrinsic dispersion of the KR is
consistent with that reported by \citet{HOS87} and \citet{KJM93} for
nearby galaxies, and is mainly due to neglecting velocity dispersions
in the FP relation (Eq.~\ref{FPEQ}).

\subsection{Comparison of the distributions of effective parameters}
\label{SDEV}
\begin{figure}
\epsscale{0.8}
\plotone{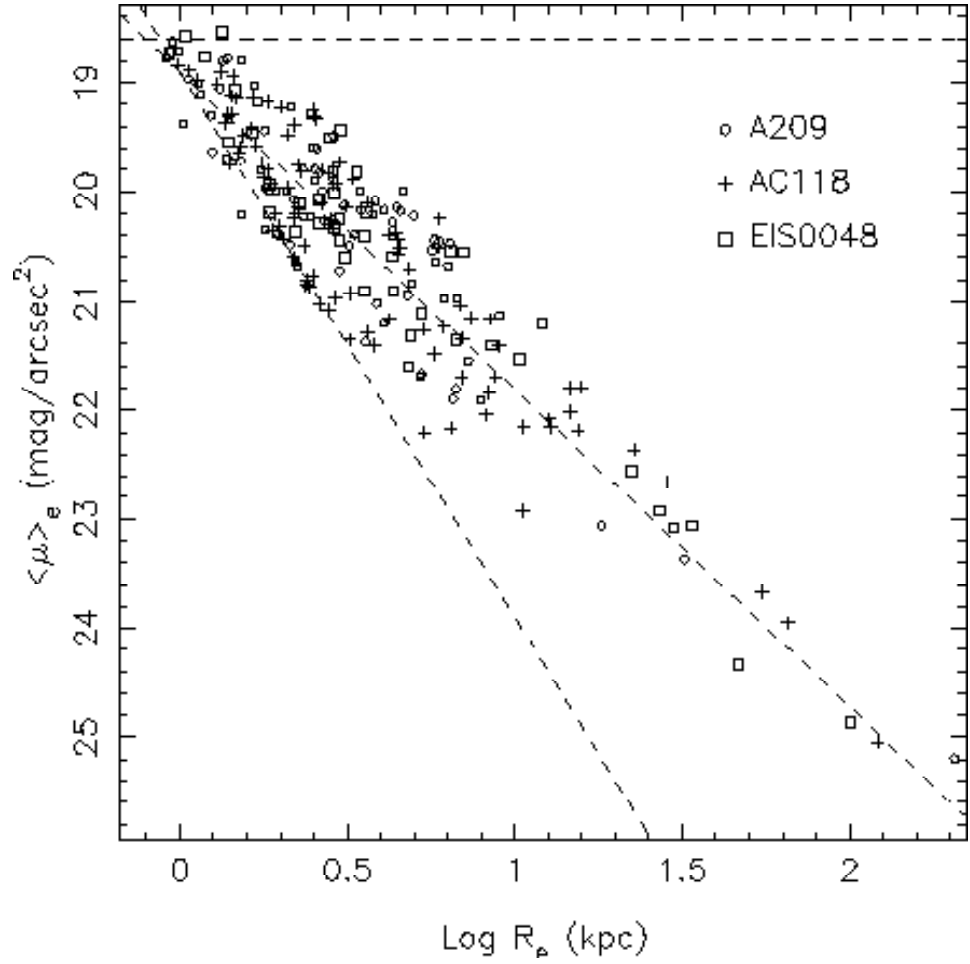}
\caption{Kormendy relations of \a209, \ac118, and \eis0048, shifted
to the same zero point.}
\label{KRTOT}
\end{figure}

To compare the distributions in the \lre--\mie \, plane, we first
shifted the values of \mie \, for \a209, \ac118, and \eis0048 by
matching the zeropoints of the KRs, and then applied to each sample
the same cuts in magnitude and \mie. Since the $\mathrm{MLS}$ fit
provides un-biased estimates of the KR coefficients, this procedure is
not affected by the different selection criteria of each sample.  The
\coma \, cluster was not considered in order to analyze only the
samples with structural parameters derived by the same procedure (see
below).  The combined \lre--\mie \, diagram for \a209, \ac118, and
\eis0048 is shown in Figure~\ref{KRTOT}.

For each cluster, we derived the mean value, $\mathrm{ < \! \log r_e \! >
}$, and the standard deviation, $\mathrm{ \sigma_{\log r_e} }$, of the \lre
\, distribution by applying the bi-weight statistics
(e.g.~\citealt{BFG90}), which has the advantage to minimize the effect
of outliers. As shown in table~\ref{LRECOF}, the values of $\mathrm{
 < \! log r_e \! > }$ and $\mathrm{ \sigma_{\log r_e} }$ are fully consistent
(within $\sim 1 \sigma$) for each pair of clusters\footnote{This
result is not affected by the difference in the measurement
uncertainties on $\mathrm{r_e}$ among the clusters. In fact, by
subtracting in quadrature the error on $\mathrm{ \log r_e }$ from
$\mathrm{ \sigma_{\log r_e} }$, we obtain the following estimates of the
standard deviations: $0.25 \pm0.04$, $0.31 \pm 0.06$ and $0.30 \pm
0.08$, for \a209, \ac118, \, and \eis0048, respectively.}. In order to
test the presence of differences in the distributions of effective
radii, we also applied the Kolmogorov--Smirnov (KS) test. The
probability that the observed samples come from the same parent
distribution turned out to be $30\%$ for \a209 \, and \ac118, $42\%$
for \a209 \, and \eis0048, and $35\%$ for \ac118 \, and \eis0048,
indicating that the distribution of galaxy sizes does not change
significantly over the explored redshift interval.

In order to verify the presence of possible differences in the
\lre--\mie \, diagram among the clusters, we used the two-dimensional
KS test (\citealt{FaF87}).  The test was repeated by using different
shifts for each cluster to take into account the uncertainties on
$\alpha$ reported in table~\ref{KRPAR}. Comparing \ac118 with the
other two clusters, we found probabilities of the KS statistics
between $10\%$ and $25\%$, while a larger probability ($\sim 50\%$)
was obtained by comparing \a209 and \eis0048. We point out that the
last result is of particular interest, since the measurement
uncertainties on the parameters of \a209 and \eis0048 are similar and
are comparable with the intrinsic dispersion of the KR.  Since the
error bars for the \ac118 data are significantly larger, we also
performed the KS test by adding further scatter to the \lre -- \mie \,
diagram of \a209 \, and \eis0048 \, in order to mimic the measurement
uncertainties on the effective parameters at $\mathrm{z \sim 0.3}$.
In this case, we obtained higher probabilities, between $50\%$ and
$80\%$, that the sample of \ac118 is drawn from the same parent
distribution of the other two clusters.

By considering also the \coma \, cluster in the comparison of the \lre
-- \mie \, distributions, we obtain $\mathrm{ < \!  \log r_e \! > =
  0.40 \pm 0.03} $ and $\mathrm{ \sigma_{\log r_e} = 0.24 \pm 0.03}$,
while the KS tests for both the one-dimensional and two-dimensional
case give probabilities between $5\%$ and $10\%$. The fact that the KS
probabilities and the value of $\sigma_{\log r_e}$ are lower with
respect to those obtained for the other distributions can be explained
by the different procedure by which the structural parameters of the
\coma \, cluster are derived, and is therefore not relevant for the
present analysis.

The previous results, together with those of Sections~\ref{SBEV}
and~\ref{SSEV}, indicate that the distribution in the \lre--\mie \,
plane for cluster galaxies does not vary significantly at least back
to $\mathrm{z\sim0.64}$.

\section{Luminosity evolution}
\label{SLEV}
The evolution of the zeropoint of the KR over redshift is determined
(1) by the variation of the distribution of galaxy sizes and of the KR
slope with look-back time, (2) by the effect of the cosmological
dimming on \mie, (3) by the luminosity evolution of galaxy stellar
populations, (4) by the values of cosmological parameters. Point (1)
is ruled out by the results of Sections~\ref{SBEV} and~\ref{SDEV}.
Since the investigation of point (4) would require a larger number of
cluster samples, we will focus the discussion on points (2) and (3).

We start to note that, as for the KR slope, the value of $\alpha$
obtained by the $\mathrm{BMLS}$ fit for the Coma sample is consistent
with that found by previous studies at $\mathrm{z \sim 0}$. For
example, if we adopt the KR slope obtained by \citet[hereafter
HOS87]{HOS87} in the $\mathrm{r}$ band for a sample of 97 bright
nearby early--types ($\mathrm{\beta=3.14}$), we obtain $\alpha=18.60$
for the \coma \, sample, which is in very good agreement with that of
HOS87 reported to our cosmology\footnote{We also took into account the
k-correction and the cosmological dimming which HOS87 removed from the
data.}: $\alpha=18.55 \pm 0.13$.

Suitable corrections were applied to the data in order to convert the
\mie \, values to the same restframe waveband (V) for each cluster. We
used the GISSEL98 synthesis code (\citealt{BrC93}) to construct galaxy
templates with different age $\mathrm{t_f}$ (referred to
$\mathrm{z=0}$), metallicity $\mathrm{Z}$ and time scale $\tau$ of
star formation. In order to consider a wide range of possible spectral
types describing the galaxies in our sample, we chose the following
range of parameters: $\mathrm{t_f\in[8,13]~Gyr}$, $\mathrm{Z \in
[0.5,1.5]Z_\odot}$ and $\mathrm{\tau\in[0.01,5]~Gyr}$. The mean values
of the derived corrections and the relative uncertainties are shown in
table~\ref{KCOR}. We notice that the uncertainties are very small
($\mathrm{\lesssim 0.03~mag}$), i.e. the dependence on the spectral
type is negligible, due to the fact that the observed wavebands
approximate closely the V-band restframe for each cluster.  We also
verified that using SSP spectra by \citet{BUZ89} gives values fully
consistent with those of table~\ref{KCOR}.  This proves that it is
correct to apply the same corrections to all the galaxies in each
cluster.  The V-band zeropoints ($<\!\mu_V\!>_e$) of the KR were
obtained by adding the values reported in table~\ref{KCOR} to the
values of $\alpha$ (table~\ref{KRPAR}), and the corresponding error
budgets were estimated by adding in quadrature the uncertainties on
the corrections with $\sigma_\alpha$.

\begin{figure}
\epsscale{0.8}
\plotone{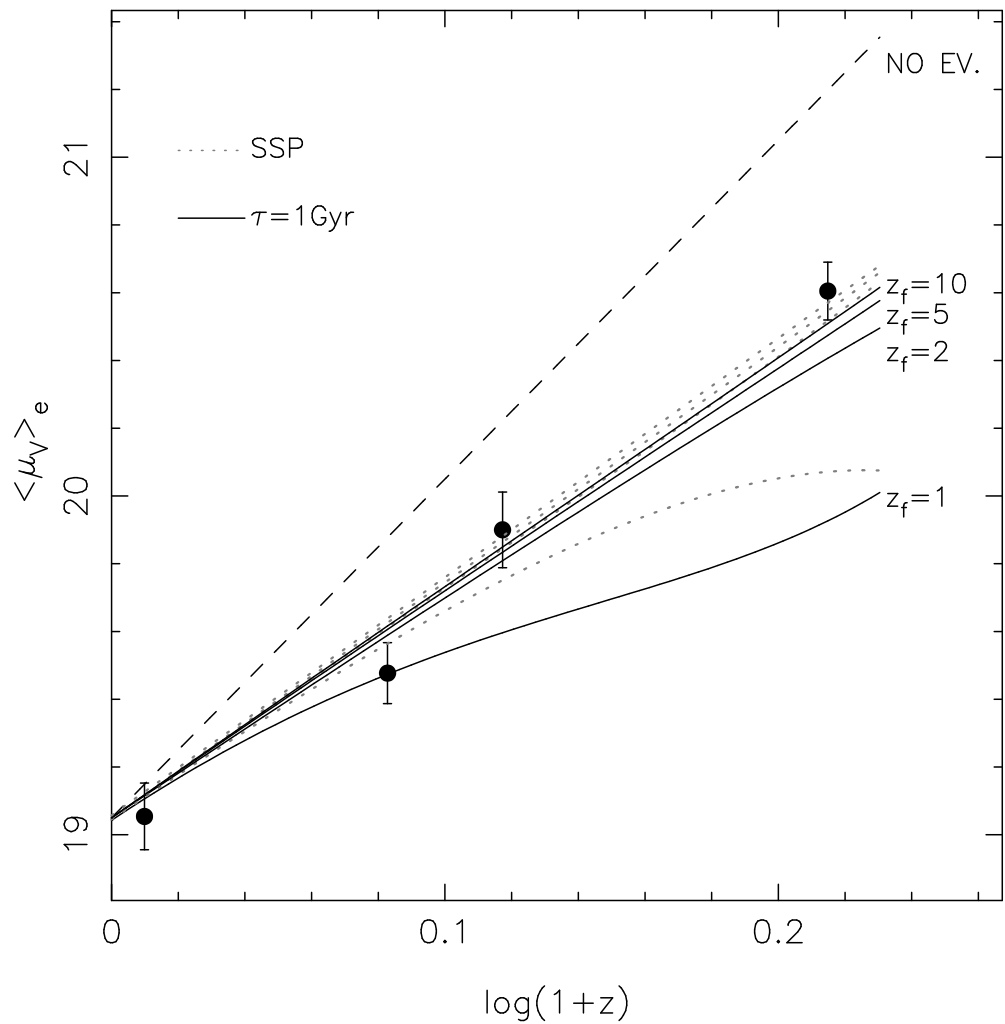}
\caption{Evolution of the zeropoint of the KR. The dashed line
indicates the expected change for pure cosmological dimming. The other
curves are $\mathrm{dimming+passive}$ evolution models corresponding
to a SSP with formation redshifts $\mathrm{z_f=1,2,5,10}$ (dotted
lines) and to a population with an exponential SFR
($\tau=\mathrm{1~Gyr}$) starting at the redshifts used for the SSP
(solid lines). }
\label{LEV}
\end{figure}

In Figure~\ref{LEV} we show the relation between $<\!\mu_V\!>_e$ and
$\mathrm{\log(1+z)}$. The observed points are well described by a
linear relation $\mathrm{<\!\mu_V\!>_e=\Gamma \cdot \log(1+z)}$. A
least square fit gives $\Gamma=7.7 \pm 0.6$, or equivalently:
$\mathrm{<\!I_V\!>_e \propto (1+z)^{-3.08\pm0.24}}$, where
$\mathrm{<\!I_V\!>_e}$ is the mean surface brightness within \re \, in
linear units. As shown by \citet{LuS01c}, this excludes the relation
$\mathrm{<\!I_V\!>_e \propto (1+z)^{-1}}$, which would hold in the
case of a non-expanding universe.  We compare the relation between
$<\!\mu_V\!>_e$ and $\mathrm{\log(1+z)}$ with a no evolution sequence,
that is with a pure Tolman signal $\mathrm{<\!I_V\!>_e \propto
(1+z)^{-4}}$, and with models that include both cosmological dimming
and passive luminosity evolution (see Figure~\ref{LEV}).  These models
were constructed by the GISSEL98 synthesis code as follows. We
considered galaxy templates with solar metallicity
$\mathrm{Z=Z_\odot}$, with different formation redshifts
($\mathrm{z_f= 1,2,5,10}$), and with two time scales of star
formation: $\mathrm{\tau=1~Gyr}$, which gives a suitable description
of the evolution of ETG integrated colours, and
$\mathrm{\tau=0.01~Gyr}$, which reproduces the properties of a simple
stellar population. Changing the value of $\mathrm{Z}$ in the range
$0.5$--$1.5Z_\odot$ does not change significantly the results.  The
models were arbitrarily scaled to fit the data.

According to what found by previous studies (see~\citealt{LuS01c}),
the trend of $<\!\mu \!>_e$ with redshift is well described by the
superposition of the Tolman signal with pure (passive) luminosity
evolution. The formation redshift of the stellar populations is
constrained to be higher than $\mathrm{z_f\sim2}$ with a confidence
level $\gtrsim2 \sigma$, in agreement with what found by previous
studies of the KR (\citealt{BAS98} and \citealt{ZSB99}) and of the FP
(\citealt{KvDF97}; \citealt{vDF98b}; \citealt{J99}; \citealt{PDdC01}).

\section{Conclusions}
\label{SCONC}
We have studied the evolution of the optical, V-band restframe,
Kormendy relation in the range $\mathrm{0 \la z \la 0.64}$, by using
ground-based effective parameters derived in previous works for the
clusters of galaxies \a209 ($\mathrm{z=0.21}$), \ac118 \,
($\mathrm{z=0.31}$), and \eis0048 \, ($\mathrm{z=0.64}$), and
published data for the Coma cluster ($\mathrm{z \sim
0.023}$). Galaxies were selected by the photometric redshift technique
for \ac118 \, and \eis0048 \, and by spectroscopic redshifts
complemented with photometric data for \a209.

We introduce a new fitting procedure, the $\mathrm{MLS}$ fit, which
provides un-biased estimates of the KR coefficients. This allows, for
the first time, to investigate the evolution of the slope and of the
scatter of the KR, as well as of the whole distribution in the plane
of effective parameters.

The results of the present work can be summarized as follows.
\begin{description}
\item[$\bullet$] Spheroids define a tight sequence in the plane of the
effective parameters at least back to $\mathrm{z \sim 0.64}$.
\item[$\bullet$] The slope $\beta$ and the intrinsic dispersion \smie
\, turn out to be consistent at all redshifts. The mean values of
$\beta$ and \smie \, are $2.92\pm0.08$ and $0.40\pm0.03$,
respectively.
\item[$\bullet$] The fact that the KR slopes are the same among all
the clusters constrains the possible change in the formation epoch of
galaxy stellar populations per decade of radius to be smaller than
$\sim 28 \%$ for SSP models, and smaller than $\sim 18 \%$ for
populations having an exponential SFR with $\mathrm{\tau=1~Gyr}$.
\item[$\bullet$] The mean value and the width of the distribution of
galaxy sizes are consistent for each pair of clusters. Moreover, the
Kolmogorov--Smirnov tests indicate that the whole distribution in the
\lre -- \mie \, plane do not vary over redshift.
\item[$\bullet$] In agreement with previous studies, the evolution of
the KR zeropoint is explained by the superposition of the Tolman
signal and the passive evolution of stellar populations with high
formation redshift, $\mathrm{z_f>2}$.
\end{description} 

These results imply that, whatever the mechanism driving galaxy
evolution is, (1) it has already built up the Kormendy relation at
$\mathrm{z=0.64}$ and (2) it does not affect significantly the
properties of spheroids in the plane of effective parameters from that
redshift down to $\mathrm{z\sim0}$.  The fact that the KR slope does
not evolve with redshift sets constrains on the differential
luminosity evolution along the early-type sequence, and, therefore, on
the properties of galaxy stellar populations as a function of galaxy
radius.  These results should be taken into account by any model aimed
at reproducing the physical processes underlying formation and
evolution of spheroids.

%Since the uncertainty on the slope of the KR is significantly lower
%with respect to previous studies, due to the large number of galaxies
%for each cluster and to the use a suitable fitting procedure.
%Finally, we point out that the use of ground-based data holds
%interesting prospects for the employment of the KR to constrain
%cosmological models.

\acknowledgments We are grateful to R. de Carvalho and A. Iovino for
the helpful discussions. We thank A. Mercurio for providing us with
the photometry of the cluster of galaxies \a209.  Michele Massarotti
is partly supported by a `MIUR-COFIN' grant.

\appendix
\section{The MLS fits}
\label{appA}
We consider the statistical model:
\begin{equation}
\mathrm{
Y = A + B \cdot X \label{A1}
} \label{LMOD}
\end{equation}
where $\mathrm{X}$ and $\mathrm{Y}$ are two random variables
describing the distributions of \lre \, and \mie \, (or viceversa) and
A and B are the zero point and the slope of the KR.  In order to
derive A and B, we have to take into account the constrains on
$\mathrm{X}$ and $\mathrm{Y}$ arising from the selection cuts. By
considering the case of a single cut, we have the following constrain:
\begin{equation}
\mathrm{
y < c_1 + c_2 \cdot x \label{A2}
} 
\end{equation}
where x and y are the outputs of $\mathrm{X}$ and $\mathrm{Y}$, while
the coefficients $\mathrm{c_1}$ and $\mathrm{c_2}$ describe the
selection cut.  The direction of the inequality in Eq.~\ref{A2} is
chosen arbitrarily.  In the case of the magnitude limit $\mathrm{m_c}$,
with $\mathrm{x=\log r_e}$ and $\mathrm{y=<\!\mu\!>_e}$, we have
$\mathrm{c_1=m_c+2.5 \cdot \log(2\pi)}$ and $\mathrm{c_2=5}$.  This
follows from the relation $\mathrm{m_T = -2.5\cdot\log(2\pi) -5}$
$\cdot \mathrm{\log r_e + <\!\mu\!>_e}$, where $\mathrm{m_T}$ is the
total magnitude, and from the constrain $\mathrm{m_T < m_c}$.

For a fixed value of $\mathrm{x}$, the probability to find a point in
the interval $\mathrm{(y,y+dy)}$ is given by
\begin{equation}
\mathrm{
P_x(y) d \! y = K(A,B,c_1,c_2) \cdot 
                exp\left[-(y-A-B\cdot x)^2/(2\sigma^2)\right] \cdot 
f(c_1+c_2 \cdot x - y) d \! y
} \label{PMOD}
\end{equation}  
where f is a step function, and the coefficient K is derived by
imposing $\mathrm{\int P(y) d \! y = 1}$. Eq.~\ref{PMOD} can be
directly generalized to the case of multiple cuts and to a more
complex shape of the selection function f.  We write the likelihood
$L\mathrm{=\prod P_{x_i}(y_i)}$, and derive the coefficients in
Eq.~\ref{LMOD} by minimizing the function $- \mathrm{ln}L$ with
respect to A and B.  We notice that in the ordinary least square fit
the function f is absent in Eq.~\ref{PMOD}. Therefore, the term K does
not depend on $\mathrm{A}$ and $\mathrm{B}$ and the minimization of $-
\mathrm{ln}L$ reduces to solving a coupled system of linear equations
in A and B.

\section{Parameterizing the luminosity evolution}
\label{appB}
Spectra of galaxy templates were constructed by the GISSEL98 synthesis
code, considering models with a Scalo IMF and an exponential SFR,
$\mathrm{e^{-t/\tau}}$. We considered SSP spectra and models having a
protracted SFR, $\tau =\mathrm{1.0~Gyr}$, with
metallicity $\mathrm{Z=0.2, 0.4, 1, 2.5~Z_\odot}$ and age $\mathrm{t_f
\in [4,13]~Gyr}$.  The range of $\mathrm{t_f}$ was chosen in order to
describe the properties of galaxy templates having formation redshift
$\mathrm{z_f \in [1.8,10]}$ (corresponding to a look-back time
$\mathrm{t_{LB} \in [10,13]~Gyr}$), from $\mathrm{z=0.64}$ to
$\mathrm{z=0}$.

For both models, we found that the dependence of the V-band magnitude
at $z\sim0$ on $\mathrm{t_f}$ and $\mathrm{Z}$ is well described by a
second-order polynomial fit:
\begin{equation}
\mathrm{ 
m_V(t_f,Z) \simeq \sum_{i=1,2} c_i \cdot \left[ \log t_f
\right]^i + \sum_{i=1,2} d_i \cdot \left[ \log Z \right]^i + cost, 
}
\label{B1}
\end{equation}
with a rms of $\mathrm{\sim 0.02~mag}$. The restframe V-band magnitude
at $\mathrm{z=0.64}$ is obtained by substituting $\mathrm{t_f}$ with
$\mathrm{t_f-t_{0.64}}$ in Eq.~\ref{B1}, where $\mathrm{t_{0.64}}$ is
the look-back time corresponding to $z=0.64$. We point out that the
presence of cross terms in Eq.~\ref{B1}, such as $\mathrm{\log t_f
\cdot \log Z}$, reduces the rms of the fit only marginally, by less
than $\mathrm{0.01~mag}$, and that, therefore, the luminosity
evolution $\mathrm{E=m_V(t_f-t_{0.64},Z) - m_V(t_f,Z)}$ depends mainly
on $\mathrm{t_f}$, that is $\mathrm{E \simeq E(t_f)}$.

The derivative of $\mathrm{E(t_f)}$ with respect to $\mathrm{t_f}$ can
be computed analytically from Eq.~\ref{B1} and is given by the
following equation:
\begin{equation}
\mathrm{ 
F(t_f)= \frac{d \! E}{d \! \log t_f} = c_1 \cdot \left[
\frac{t_f}{t_f-t_{0.64}} - 1 \right] + 2 c_2 \cdot \left[ \frac{t_f
\cdot \log (t_f-t_{0.64})}{t_f-t_{0.64}} - \log t_f \right] 
}.
\end{equation}
This relation was used to compute the range of values of
$\mathrm{F(t_f)}$ in section~\ref{SBEV}.

\begin{deluxetable}{lcccccc}
\tabletypesize{\small} 
\tablecaption{Samples used in the present study. \label{DATA} Column
1: cluster identification. Column 2: redshifts. Column 3:
waveband. Column 4: sample size. Columns 5, 6: limits in \mie \, and
magnitude. Column 7: references for data reduction and derivation of
structural parameters with the following abbreviations: \citet{JFK95a,
JFK95b, J99, LBM02, BML02, LMI03a, LBM03b, MGB03a, MMM03b} (JFK95a,b; J99; 
LBM02; BML02; LMI03; LBM03; MGB03; MMM03).}
\tablewidth{0pt}
\tablehead{ 
 & \colhead{z} & \colhead{Waveband} & \colhead{N} & \colhead{$\mathrm{<\!\mu\!>_{e,c}}$} & \colhead{$\mathrm{mag}$} & \colhead{Ref.} \\
}
\startdata
 COMA    & 0.023 & $\mathrm{ r_g }$ &  90 &   ...  &  15.3  & JFK95a, JFK95b, J99 \\
\a209    & 0.21  & $\mathrm{ R }$   &  81 &  18.6  &  20.1  & MGB03, MMM03, LBM03  \\
\ac118   & 0.31  & $\mathrm{ R }$   & 101 &  19.1  &  20.8  & BML02, LBM02  \\
\eis0048 & 0.64  & $\mathrm{ I }$   &  46 &  18.9  &  22.0  & LMI03, LBM03  \\
\enddata
\end{deluxetable}
\begin{deluxetable}{l|ccc}
\tabletypesize{\small} \tablecaption{Slopes of the KR for COMA, 
\a209, \ac118, and \eis0048.  Columns 2, 3 and 4: values 
obtained by the different fitting procedures (see text),
with the relative uncertainties ($1 \sigma$ standard
intervals). \label{KRSLOPE}} \tablewidth{0pt} \tablehead{ &
\colhead{$\mathrm{MLS_{\log r_e}}$} & \colhead{$\mathrm{MLS_{< \! \mu
\! >_e}}$} & \colhead{$\mathrm{BMLS}$} \\ } 
%% 3.9 \pm 0.4    2.5 \pm 0.3   3.01 0.26
\startdata 
$\mathrm{COMA }$ & $3.9 \pm 0.4 $ & $2.5 \pm 0.3$ & $3.01 \pm 0.26 $ \\ 
\a209  & $3.04 \pm 0.25$ & $2.70 \pm 0.17$ & $2.86 \pm 0.17$ \\ 
\ac118 & $2.84 \pm 0.20$ & $2.67 \pm 0.15$ & $2.74 \pm 0.16$ \\ 
\eis0048 & $3.11 \pm 0.13$ & $2.97 \pm 0.13$ & $3.04 \pm 0.13$ \\ 
\enddata
\end{deluxetable}
%%  3.90182  0.36234  2.36520  0.29092  2.95724  0.27932
\begin{deluxetable}{l|cc}
\tabletypesize{\small}
\tablecaption{Mean values of the slope $\beta$ of the KR obtained by the
different fitting procedures. The uncertainties $\sigma_\beta$
denote $1 \sigma$ standard intervals. \label{KRMSLOPE}}
\tablewidth{0pt}
\tablehead{  & \colhead{$\beta$} & \colhead{$\mathrm{\sigma_{\beta}}$} \\
}
\startdata
$\mathrm{MLS_{\log r_e}}$        & 3.08 & 0.10 \\
$\mathrm{MLS_{< \! \mu \! >_e}}$ & 2.78 & 0.08 \\
$\mathrm{BMLS}$                  & 2.92 & 0.08 \\
\enddata
\end{deluxetable}
\begin{deluxetable}{l|ccc}
\tabletypesize{\small} \tablecaption{Zeropoint and scatter of the
KR. Uncertainties denote $1 \sigma$ standard
intervals. $\sigma_{<\!\mu\!>_e}$ is the observed scatter in \mie,
while \smie \, is the estimate of the intrinsic dispersion of the KR.
\label{KRPAR}
}
\tablewidth{0pt}
\tablehead{ & $\alpha$ & $\sigma_{<\!\mu\!>_e}$ & \smie \\
}
\startdata
$\mathrm{COMA }$           & $18.68 \pm 0.08$ & $0.45 \pm 0.05$ & $0.43\pm0.05$\\
\a209                      & $18.95 \pm 0.08$ & $0.51 \pm 0.04$ & $0.36\pm0.07$\\
\ac118                     & $19.45 \pm 0.11$ & $0.60 \pm 0.05$ & $0.39\pm0.08$\\
\eis0048                   & $19.51 \pm 0.07$ & $0.42 \pm 0.04$ & $0.40\pm0.04$\\
\enddata
\end{deluxetable}
\begin{deluxetable}{lcc}
\tabletypesize{\small} \tablecaption{Mean value, $\mathrm{ < \! log
r_e \! > }$, and standard deviation, $\mathrm{ \sigma_{\log r_e} }$,
of the \lre \, distributions. The uncertainties, which indicate $1
\sigma$ standard intervals, were obtained by the bootstrap method.
\label{LRECOF}
}
\tablewidth{0pt}
\tablehead{ 
 & \colhead{$\mathrm{ < \! \log r_e \! > } $} & \colhead{$\mathrm{ \sigma_{\log r_e} }$} \\
}
\startdata
\a209     & $0.50 \pm 0.06$ & $0.27 \pm 0.04$ \\
\ac118    & $0.46 \pm 0.05$ & $0.34 \pm 0.05$ \\
\eis0048  & $0.48 \pm 0.03$ & $0.31 \pm 0.08$ \\
\enddata
\end{deluxetable}
\begin{deluxetable}{l|c}
\tabletypesize{\small}
\tablecaption{Correction to V-band restframe for the cluster samples.
Uncertainties denote $1 \sigma$ standard intervals. \label{KCOR}
}
\tablewidth{0pt}
\tablehead{ & \colhead{$\mathrm{COR.}$}\\
}
\startdata
$\mathrm{COMA \, (JFK96)}$ & $0.373 \pm 0.040$ \\
\a209                      & $0.527 \pm 0.004$ \\
\ac118                     & $0.450 \pm 0.020$ \\
\eis0048                   & $1.095 \pm 0.030$ \\
\enddata
\end{deluxetable}

\newpage


\end{document}